\documentstyle[12pt]{article}

\global\arraycolsep=1pt
\begin{document}

\hfill    SISSA/ISAS 66/94/EP

\hfill    IFUM 470/FT

\hfill    hepth@xxx/9405175

\hfill    May, 1994

\begin{center}
\vspace{10pt}
{\large \bf
FROM N=2 SUPERGRAVITY TO CONSTRAINED MODULI SPACES
\, \footnotemark\footnotetext{Talk given at the Trieste Workshop on
String Theory, April 1994.}
\footnotemark\footnotetext{
Partially supported by EEC, Science Project SC1$^{*}$-CT92-0789.}}
\vspace{10pt}

{\sl Damiano Anselmi, Pietro Fr\'e}

\vspace{4pt}

International School for Advanced Studies (ISAS), via Beirut 2-4,
I-34100 Trieste, Italia\\
and Istituto Nazionale di Fisica Nucleare (INFN) - Sezione di Trieste,
Trieste, Italia\\

\vspace{8pt}

{\sl Luciano Girardello and Paolo Soriani}

\vspace{4pt}

Dipartimento di Fisica, Universit\`a di Milano, via Celoria 6,
I-20133 Milano, Italia\\
and Istituto Nazionale di Fisica Nucleare (INFN) - Sezione di Milano,
Milano, Italia\\
\end{center}

\vspace{8pt}

\begin{center}
{\bf Abstract}
\end{center}

\vspace{4pt}
\noindent

In this talk we review some results concerning
a mechanism for reducing the moduli space of a topological
field theory to a proper submanifold of the ordinary moduli space.
Such mechanism is explicitly realized in the example of
constrained topological gravity, obtained by topologically twisting
the N=2 Liouville theory.

\vfill
\eject

\newcommand{\be}{\begin{equation}}
\newcommand{\ee}{\end{equation}}
\newcommand{\ba}{\begin{eqnarray}}
\newcommand{\ea}{\end{eqnarray}}
\newcommand{\ban}{\begin{eqnarray*}}
\newcommand{\ean}{\end{eqnarray*}}
\newcommand{\brr}{\begin{array}}
\newcommand{\err}{\end{array}}
\newcommand{\bc}{\begin{center}}
\newcommand{\ec}{\end{center}}
\newcommand{\sss}{\scriptscriptstyle}
\newcommand{\bea}{\begin{eqnarray}}
\newcommand{\eea}{\end{eqnarray}}
\newcommand{\bean}{\begin{eqnarray*}}
\newcommand{\eean}{\end{eqnarray*}}

\def\bz{{\bar z}}
\def\s{s^\prime}
\def\t#1{\tilde #1}
\def\La{\Lambda}
\def\lamb{\lambda}
\def\o#1#2{{{#1}\over{#2}}}
\def\nn{\nonumber}
\def\is{{I^*}}
\def\js{{J^*}}

Topological field theories \cite{witten} represent an amazing joint-venture
between mathematics and physics. They can be divided, according to Witten
,
in two broad classes:
the {\sl cohomological, or semiclassical theories}, whose
prototypes are either the
topological Yang-Mills theory \cite{wittenym} or the
topological $\sigma$-model \cite{wittensigma}
and the  {\sl quantum theories}, whose prototype is
the abelian Chern-Simons theory
\cite{chernsimons}.

In this talk we are concerned
with cohomological theories. The basic idea is that
a generic correlation function of $n$ physical observables
$\{ {\cal O}_1 , \ldots, {\cal O}_n \}$
has an interpretation as the {\sl intersection number}
\begin{equation}
<{\cal O}_1  {\cal O}_2 \cdots {\cal O}_n > \,=\,\# \left (  H_1  \cap  H_2
 \cap  \cdots \cap H_n  \right )
\label{intro1}
\end{equation}
of $n$ {\sl homology cycles} $H_i \, \subset \, {\cal M}$ in the {\sl moduli
space} ${\cal M}$
of suitable {\sl instanton} configurations $\Im \left [\phi (x) \right ]$ of
the basic
fields $\phi$ of the theory.

Topological field theories can been defined in completely geometrical terms.
However, in every topological model,
the right hand side of equation (\ref{intro1}) should admit an
independent definition as a
functional integral in a suitable Lagrangian quantum field theory,
in order to be of physical interest.

We present the idea of
reducing the moduli space to a constrained submanifold
of the ordinary moduli space and analyze the field
theoretical mechanism that implements such a reduction.
The results are based on ref.\ \cite{preprint}.
The specific model that suggested this idea to us
comes from the twist of N=2 Liouville theory and it
is called by us
{\sl constrained topological gravity}. The physical correlators are
{\sl intersection numbers} in a proper submanifold ${\cal V}_{g,s}
\subset {\cal M}_{g,s}$
of the moduli space ${\cal M}_{g,s}$ of genus $g$ Riemann surfaces
$\Sigma_{g,s}$ with $s$ marked points.
${\cal V}_{g,s}$ is defined as follows.
Consider the $g$-dimensional vector bundle
${\cal E}_{hol} \, \longrightarrow \, {\cal M}_{g,s}$, whose
sections $s(m)$ are the holomorphic
differentials $\omega$ on the Riemann surfaces
$\Sigma_{g,s}$, $m$
denoting the point of the base-manifold ${\cal M}_{g,s}$
(i.e.\ the {\sl polarized} Riemann surface).
Let $c({\cal E}_{hol})=\det (1+{\cal R})$ be the total Chern class
of ${\cal E}_{hol}$, ${\cal R}$
being the curvature two-form of a holomorphic connection on
${\cal E}_{hol}$.
Then ${\cal V}_{g,s}$ is the Poincar\'e dual
of the top Chern class $c_g({\cal E}_{hol})=\det {\cal R}$.
${\cal V}_{g,s}$ is a submanifold
of codimension $g$ which can be described as the locus of
those Riemann surfaces $\Sigma_{g,s}(m)$ where
some section $s(m)$ of ${\cal E}_{hol}$ vanishes \cite{griffithsharris}.

Explicitly, the
topological correlators of constrained topological gravity
are the intersection numbers of the standard Mumford-Morita
cohomology classes $ c_1 \left ( {\cal L}_i \right )$ on the constrained
moduli space, namely
\begin{eqnarray}
<{\cal O}_1 \left ( x_1 \right )  {\cal O}_2 \left ( x_2 \right )
&\cdots &{\cal O}_n \left ( x_n \right )  > \, =\int_{{\cal V}_{g,s}}
\left [  c_1  \left ( {\cal L}_1 \right ) \right ]^{d_1}  \wedge  \cdots\wedge
\left [  c_1  \left ( {\cal L}_n \right ) \right ]^{d_n}=\nonumber\\&
=&\int_{{{\cal M}}_{g,s}} c_g (  {\cal E}_{hol})\wedge
\left [  c_1 \left ( {\cal L}_1 \right ) \right ]^{d_1}  \wedge \cdots
\wedge\left [  c_1 \left ( {\cal L}_n \right ) \right ]^{d_n}.
\label{intro_0}
\end{eqnarray}
Precisely,
$ c_1 \left ( {\cal L}_i \right )$ are the first Chern-classes
of the bundles
${\cal L}_i\longrightarrow{\cal M}_{g,s}$ whose sections
are elements of the form $h(m)dz_i$ of the cotangent bundle
$T^{ * }_{x_i} \Sigma_{g}(m)$ at the marked point $x_i=(z_i,\bar z_i)$.

The origin of a constraint on moduli space is due to the presence
of the graviphoton in the N=2 graviton multiplet.
The graviphoton is initially a physical gauge-field and
after the twist maintains zero ghost-number. Nevertheless,
in the twisted theory, it is no longer a physical field,
rather it is a Lagrange multiplier (in the BRST sense). Indeed,
it appears in the right-hand side of the BRST-variation of suitable
antighosts, coming from some components of the
gravitini. Since this Lagrange multiplier possesses global degrees
of freedom (the $g$ moduli of the graviphoton), it
imposes $g$ constraints on the space ${\cal M}_{g}$,
which is the space
of the global degrees of freedom of the metric tensor. The metric
tensor,
on the other hand, is the only
field that remains physical also after twist.
We are lead to conjecture that
the inclusion of {Lagrange multiplier gauge-fields}
is a general mechanism producing constraints on the
moduli spaces.

The gauge-free BRST algebra ${\cal B}_{gauge-free}$ is the same as
in the Verlinde and Verlinde model \cite{verlindesquare},
based on the gauge group $SL(2,{\bf R})$.
The flat $SL(2,R)$ connection
$\left \{ e^\pm , e^0 \right \}$
contains the zweibein $e^\pm$
and the spin connection of a constant curvature metric on the imaginary
upper half-plane $H$. The BRST quantization of the most general
continuous deformation of the $SL(2,R)$ connection is derived
in the standard way. The (off-shell) gauge-fixing BRST algebra
${\cal B}_{gauge-fixing}$, on the other hand, is of the following type
\begin{equation}
s{\bar\psi}=A-d{\gamma},\quad\quad sA=-dc,\quad\quad s\gamma=c,\quad
\quad sc=0,
\label{intro2}
\end{equation}
where $A$ is the graviphoton,
${\bar \psi}$ is a one-form of ghost number $-1$, coming from the gravitini,
${\gamma}$ is a zero-form of ghost number $0$ and
$c$ is the ordinary gauge ghost (with ghost number $1$).

The true (complex)
dimension of the subspace ${\cal V}_{g,s}$ is ${\rm dim}_{\bf
C}\,{\cal V}_{g,s}=2g-3+s$, so that the selection rule for (\ref{intro1}) to
be nonvanishing is
\begin{equation}
\sum_id_i=2g-3+s,
\end{equation}
or
\begin{equation}
\sum_i(d_i-1)=2g-3={\rm dim}_{\bf C}\, {\cal V}_g.
\end{equation}
However, the formal (real) dimension of the moduli space turns out to be
$4g-4$, instead of $4g-6$, so that one has to satisfy
\begin{equation}
\sum_ig_i=4g-4,
\end{equation}
$g_i$ being the ghost number of ${\cal O}_i$. The fact that the true
dimension
is smaller than
the formal dimension is only apparently puzzling and can be understood
as follows.
Antighost zero-modes correspond
to local vector fields normal to the constrained surface and ghost
zero-modes correspond to possible obstructions to the
globalization of such local vector fields. As a consequence, the difference,
in the constraint sector of the BRST algebra, of antighost zero-modes
minus ghost zero-modes,  expresses the minimum number of
constraints that are imposed. If the potential obstructions do not occur,
then all the
antighosts correspond to actual normal directions to the constrained
surface and the true
dimension of the constraint surface is smaller than its formal dimension.
As one sees, rules and roles are reversed with respect to the
ordinary case.

At the level of conformal
field theories there is also a crucial difference between constrained
and ordinary topological gravity, which
keeps trace of the constraint on moduli space.
Indeed, after gauge-fixing
and in the limit where the cosmological constant
tends to zero, our model also reduces to the sum
of two topological
conformal field theories $Liouville\, \oplus \, Ghost$;
the central charges, however, are $c_{Liouville}=6$
and  $c_{Ghost}=-6$,
rather than $3$ and $-9$.

\section{N=2 D=2 supergravity and its twist}
\label{N2D2}

We assume that the Lagrangian of N=2 supergravity
is the supersymmetrization of the following Largangian of pure gravity,
\begin{equation}
{\cal L}_{Liouville} = \Phi(R[g]+a^2)\sqrt { \det  g},
\label{intro924}
\end{equation}
$\Phi$ being an independent field (the dilaton).
The result is
\begin{equation}
{\cal L}={\cal L}_1+{\cal L}_2,
\end{equation}
where ${\cal L}_1$ and ${\cal L}_2$ are the kinetic and de Sitter
terms, respectively,
\begin{eqnarray}
{\cal L}_1 &=& (X + \bar X)R - \o{i}{2}(X - \bar X) F
- 2 \lamb_- \rho^- + 2 \lamb_+ \rho^+
+ 2 \tilde\lamb^- \tilde\rho_- - 2\tilde\lamb^+
\tilde\rho_+ \nonumber\\&&
- 4i \bar M H e^+ e^- + 4i M \bar H e^+ e^-,\nonumber\\
{\cal L}_2 &=& (MX + \bar M\bar X) e^+ e^-
+\lamb_-  \tilde\zeta_+ e^+ - \lamb_+ \tilde\zeta_- e^+ +  \tilde\lamb^-
\zeta^+ e^-  - \tilde
\lamb^+
\zeta^- e^- \nn\\
&+& {i\over 2} X
\zeta^+ \tilde\zeta_+ + {i\over 2}\bar X\zeta^- \tilde\zeta_-
+ 2i (\bar H -H)e^+ e^-.
\label{lagra}
\end{eqnarray}
$\{e^\pm,\zeta^\pm,\tilde\zeta_\pm,A,M,\bar M\}$ is the graviton
multiplet, $M$ and $\bar M$ being auxiliary fields, and
$\{X,\bar X,\lambda_\pm,\tilde\lambda^\pm,H,\bar H\}$ is the
dilaton multilet, $H$ and $\bar H$ being auxiliary fields.
It is easy to see that using the equation of motions of
$H$, $\bar H$ and $X+\bar X$,
we get precisely a de Sitter supergravity with
cosmological constant $\Lambda=\o{1}{2}$. The field strength $F$, on
the other hand, is set to zero by the $X-\bar X$ field equation.

The superalgebra of the graviton multiplet possesses some interesting features
that we do not think have been previously noticed in the literature and
that are related to the peculiar properties of the graviphoton after twist.
In particular, one can show that the off-shell supersymmetry transformations
can be found if and only if the gravitini are $U(1)$ charged.
That is why there is a nontrivial $U(1)$ current. After twist, this current
is viewed as a section of ${\cal E}_{hol}$ and is set to zero by the
global degrees of freedom of the Lagrange multiplier $A$,
thus realizing
the projection onto the Poincar\'e dual of $c_g({\cal E}_{hol})$.

Technically, the A and B twists are performed as follows.
To begin with, we have to
notice that the Lagrangian ${\cal L}_1$ of Poincar\'e  gravity,
possesses a global $R$-symmetry
[which will be denoted by $U(1)^\prime$], under which
the fields transform with the following charges: $\zeta^\pm$,
$\tilde\zeta_\pm$, $\lambda_\mp$ and $\tilde\lambda^\mp$ have
charge $\pm 1/2$; $M$ and $H$ have charge $1$, while $\bar M$ and
$\bar H$ have charge $-1$.
$U(1)^\prime$ is not a local symmetry and it is not even a global
symmetry for the de Sitter Lagrangian ${\cal L}_2$ (\ref{lagra}).
Depending on the choice of the twist (A or B), the new Lorentz group is
defined as a combination of the old one with the $U(1)^\prime$
or $U(1)$ symmetry; viceversa for the ghost number.

We focus here on the A twist.
The new assignments  and the topological shift are
\begin{eqnarray}
\begin{array}{ll}
{\rm spin}^\prime = \hbox{spin} + U(1)^\prime, \quad &\quad
\Gamma^+ \to  \Gamma^+ + \alpha,\\
{\rm ghost}^\prime = {\rm ghost} + 2 U_A (1),\quad &\quad
\tilde\Gamma_- \to  \tilde\Gamma_- +\beta,
\end{array}
\label{spina}
\end{eqnarray}
$\Gamma^+$ and $\tilde\Gamma_-$ are the supersymmetry ghosts associated to
the gravitini $\zeta^+$ and $\tilde\zeta_-$, respectively.
$\alpha$ and $\beta$ are the so-called {\sl brokers} \cite{ansfre}.
They are to be treated formally as constant ($d\alpha=d\beta=0$) and
their (purely formal) role is to bring the correct contributions
of spin and ghost number to the fields.

The shift produces a new BRST operator $s^\prime$ which
equals $s+\delta_T$, $\delta_T$ being the topological variation
(known as ${\cal Q}_s$ in conformal field theory) and $s$ is the
initial BRST operator. On the
graviton multiplet, $\delta_T$ acts as
\begin{eqnarray}
\begin{array}{ll}
\delta_T e^+ ={i\over 2}\alpha \zeta^- \quad & \quad
\delta_T e^- ={i\over 2} \tilde \zeta_+ \beta \\
\delta_T\zeta^+ = - {1\over 2} \omega \alpha - {i\over 4} A \alpha - M
\beta e^+ \equiv B_1\alpha
\quad & \quad
\delta_T \zeta^- =0\\
\delta_T \tilde\zeta_+ =0
\quad & \quad
\delta_T \tilde\zeta_- =  {1\over 2} \omega \beta - {i\over 4}
A \beta + \bar M \alpha e^- \equiv B_2\beta\\
\delta_T M=-{i\over 2}\tilde\tau_+\alpha \quad &\quad
\delta_T \bar M=-{i\over 2}\tau^-\beta
\end{array}\nonumber
\end{eqnarray}
\begin{eqnarray}
\delta_T\omega &=& {i\over 2}M\zeta^-\beta +
{i\over 2}\bar M\alpha \tilde\zeta_+ +{i\over 2}e^-\tau^-\alpha
+{i\over 2}e^+\tilde\tau_+\beta\nonumber\\
\delta_T A &=&M\zeta^-\beta -\bar M\alpha
\tilde\zeta_+ +\tau^-\alpha e^-
-\tilde\tau_+\beta e^+.
\label{deltaT}
\end{eqnarray}
Taking into account that the BRST algebra closes off-shell, we see that $B_1$
and $B_2$ play the role of Lagrange multipliers, since they are
the BRST variations of the antighosts $\zeta^+$ and $\tilde\zeta_-$.
$B_1$ and $B_2$ can be considered as redefinitions of $A$, $M$ and
$\bar M$. Indeed, since $M$ and $\bar M$ have spin $1$ and $-1$ after the
twist,
$Me^++\bar M e^-$ can be considered as a one form. In particular,
we have shown that the graviphoton $A$ belongs to
${\cal B}_{gauge-fixing}$. On the other hand, it is clear that
the gauge-free topological algebra is that of $SL(2,{\bf R})$,
since the above formul\ae\ show that the topological symmetry
is the most continuous
deformation of the zweibein.

On the dilaton multiplet $\delta_T$ is
\begin{eqnarray}
\begin{array}{ll}
\delta_T X = \tilde\lambda^- \beta \quad & \quad
\delta_T {\bar X} =  - \lambda_+ \alpha \\
\delta_T \lambda_- = -{i\over 2} \nabla_+ X \alpha + H \beta
\equiv H_1\alpha\quad & \quad
\delta_T \tilde\lambda^- = 0 \\
\delta_T \lambda_+ = 0 \quad & \quad\delta_T \tilde\lambda^+ =
{i\over 2} \nabla_- {\bar X} \beta - \bar H \equiv H_2\beta \\
\delta_T H ={i\over 2} \nabla_+ \tilde\lambda^- \alpha \quad & \quad
\delta_T {\bar H} = - {i\over 2} \nabla_- \lambda_+ \beta\\
\end{array}
\label{deltaT2}
\end{eqnarray}
$H_1$ and $H_2$ are also Lagrange multipliers, redefinitions
of $H$ and $\bar H$.

Finally, the topological variation of the brokers vanishes, but
nilpotence of $s^\prime$ and $s$ requires
\begin{eqnarray}
\begin{array}{ll}
s^\prime \alpha =- {1\over 2} C^0 \alpha - {i\over 4} C \alpha=s\alpha \quad &
\quad
s^\prime \beta =  {1\over 2} C^0 \beta - {i\over 4} C \beta=s\beta.
\end{array}
\end{eqnarray}
In other words, even if formally, $\alpha$ and $\beta$ have to be considered
as sections with definite spin and $U(1)$ charge.

Using the above formulae, one can write the
full Lagrangian $\cal L$ as the topological variation of a suitable
gauge fermion $\Psi$ plus a total derivative term.

The observables of the topological theory are easily derived, as in
the case of the Verlinde and Verlinde model, from the descent
equations $\hat d \hat R^n=0$, $\hat R=R+\psi_0+\gamma_0$ being
the BRST extension of the curvature $R$. In particular, the local
observables are
\begin{equation}
\sigma^{(0)}_n(x)=\gamma_0^n(x).
\end{equation}
On the other hand, the field
strength $F$ does not provide any new observables, due to the fact
that $A\in {\cal B}_{gauge-fixing}$.

\section{The conformal field theory associated with N=2 Liouville gravity}
\label{conformal}

Diffeomorphism are fixed by the usual conformal gauge condition:
\begin{equation}
\matrix{e^+={\rm e}^{\varphi(z,\bar z)} dz,&e^-=
{\rm e}^{\varphi(z,\bar z)}d\bar z,}
\label{gfcond1}
\end{equation}
where $\varphi(z,\bar z)$ is the conformal factor, which is to be
identified with the Liouville quantum field.

Supersymmetries are fixed by extending the conformal gauge by means of
the conditions
\begin{equation}
\matrix{\zeta^+=\eta^+_z {\rm e}^\varphi dz,&
\zeta^-=\eta^-_z  {\rm e}^\varphi dz,&
\tilde\zeta_+=\eta_{+\bar z} {\rm e}^\varphi d\bar z,&
\tilde\zeta_-=\eta_{-\bar z} {\rm e}^\varphi d\bar z,}
\label{gfcond2}
\end{equation}
where $\eta^\pm_z(z,\bar z)$ and $\eta_{\bar z}^\pm(z,\bar z)$
are anticommuting fields of spin $1/2$ and $-1/2$ (the superpartners
of the Liouville field $\varphi$).

The $U(1)$ gauge transformations have to be treated carefully, in
order to reach a complete chiral factorization
into two superconformal field theories (left and right moving).
This is because
the theory that we are now dealing with
possesses a single local $U(1)$ symmetry, the $U(1)^\prime$ R-symmetry
being only global.
Let us introduce an additional trivial BRST system
(a ``one dimensional topological $\sigma$-model'')
$\{\xi,C^\prime\}$, $\xi$ being a ghost number zero scalar
and $C^\prime$ being a ghost number one scalar.
Their BRST algebra is chosen to be
trivial, namely
\begin{equation}
\matrix{s\xi=C^\prime,&sC^\prime=0.}
\label{trivial}
\end{equation}
The meaning of this BRST system is the gauging
of the R-symmetry $U(1)^\prime$.
Indeed, $U(1)^\prime$, which is only a {\sl global} symmetry of the
starting theory, becomes a {\sl local} symmetry
in the gauge-fixed version of the same theory.
We fix both the $U(1)$ gauge symmetry and the trivial
symmetry (\ref{trivial}) by choosing the following two gauge-fixings
\begin{equation}
\matrix{A_z-\partial_{z}\xi=0,&A_{\bar z}+\partial_{\bar z}\xi=0.}
\label{1.7}
\end{equation}
corresponding to $A=*d\xi$, where $A=A_zdz+A_{\bar z}d\bar z$.

After setting $\pi=1/2 \, (X+\bar X)$ and
$\chi=i/2 \, (X-\bar X)$, and performing suitable redefinitions,
the total gauge--fixed Poincar\`e Lagrangian
takes the form
\begin{eqnarray}
{\cal L}_{Poincar\grave e}&=&
-\pi\partial_{z}\partial_{\bar z}\varphi+\chi\partial_{z}\partial_{\bar z}\xi+
\lambda_-\partial_{\bar z}\eta^-_z-\lambda_+\partial_{\bar z}\eta^+_z
\nonumber\\&&+\tilde\lambda^-\partial_{z}\eta_{-\bar z}-\tilde
\lambda^+\partial_{z}\eta_{+\bar z}-b_{zz}\partial_{\bar z} c^z
+b_{\bar z\bar z}\partial_{z} c^{\bar z}\nonumber\\&&
-\beta_{+z}\partial_{\bar z}\gamma^+-\beta_{-z}\partial_{\bar z}\gamma^-
+\beta_{+\bar z}\partial_{z}\tilde\gamma_++
\beta_{-\bar z}\partial_{z}\tilde\gamma_-
-b_z\partial_{\bar z} c+b_{\bar z} \partial_{z} \bar c.
\label{poinc}
\end{eqnarray}

It is natural to conjecture that
Poincar\'e N=2 supergravity corresponds to an N=2 superconformal
field theory. We derive
the energy-momentum tensor $T_{zz}$, the supercurrents
$G_{+z}$ and $G_{-z}$ and the $U(1)$ current $J_z$,
by first computing the BRST charge
${\cal Q}^{BRST}=\oint J_z^{BRST}dz$,
$J^{BRST}_z$ denoting the BRST current.
Acting with ${\cal Q}^{BRST}$ on the various antighost fields
it is then simple to get the ``gauge-fixings'', which are, in our
case, the N=2 currents.
We expect to have, on shell
and up to total derivative terms,
\begin{equation}
J_z^{BRST}=-c^z{\cal T}_{zz}
+{1\over 2}c{\cal J}_{z}
+{1\over 2}\gamma^+{\cal G}_{+z}-{1\over 2}\gamma^-
{\cal G}_{-z},
\end{equation}
where
\begin{equation}
\matrix{
{\cal T}_{zz}=T_{zz}^{grav}+{1\over 2}T_{zz}^{gh},&
{\cal J}_{z}=J_{z}^{grav}+{1\over 2}J_{z}^{gh},\cr
{\cal G}_{+z}={G}_{+z}^{grav}+{1\over 2}{G}_{+z}^{gh},&
{\cal G}_{-z}=G_{-z}^{grav}+{1\over 2}G_{-z}^{gh}.}
\end{equation}
The N=2 currents for the Liouville sector are
\begin{eqnarray}
T_{zz}^{grav}&=&-\partial_{z} \pi\partial_{z}\varphi+
{1\over 2}\partial_{z}^2\pi+\partial_{z}
\chi\partial_{z}\xi+{1\over 2}(\partial_{z}\lambda_-\eta^-_z-
\lambda_-\partial_{z}\eta^-_z)+
{1\over 2}(\lambda_+\partial_{z}\eta^+_z-\partial_{z}\lambda_+
\eta^+_z),\nonumber\\
G_{+z}^{grav}&=&\partial_{z}\lambda_+-\lambda_+\partial_{z}(\varphi+\xi)
+\eta^-_z\partial_{z}(\chi+\pi),\nonumber\\
G_{-z}^{grav}&=&\partial_{z}\lambda_--\lambda_-\partial_{z}(\varphi-\xi)
+\eta^+_z\partial_{z}(\chi-\pi),\nonumber\\
J_{z}^{grav}&=&\partial_{z}\chi-\lambda_-\eta^-_z-\lambda_+\eta^+_z.
\label{6.30}
\end{eqnarray}
It is easy to
check that the N=2 operator product expansions are indeed satisfied by
(\ref{6.30}), with central charge $c_{grav}=6$.

Finally the ghost currents are:
\begin{eqnarray}
T_{zz}^{gh}&=&2b_{zz}\partial_{z} c^z +\partial_{z} b_{zz} c^z +{3\over 2}
\beta_{+z}\partial_{z}\gamma^++{1\over 2}\partial_{z}\beta_{+z}\gamma^++
{3\over 2}
\beta_{-z}\partial_{z}\gamma^-+{1\over 2}\partial_{z}\beta_{-z}\gamma^-+
b_z\partial_{z} c,\nonumber\\
G_{+z}^{gh}&=&3\beta_{+z}\partial_{z} c^z +2\partial_{z}\beta_{+z} c^z
-\gamma^-b_{zz}- \gamma^-\partial_{z} b_z-2  \partial_{z}\gamma^-b_z
-\beta_{+z} c,
\nonumber\\
G_{-z}^{gh}&=&3\partial_{z} c^z \beta_{-z}+2 c^z \partial_{z}\beta_{-z}
-b_{zz}\gamma^++\partial_{z} b_z \gamma^++2 b_z \partial_{z}
\gamma^++c\beta_{-z},
\nonumber\\
J_{z}^{gh}&=&\beta_{-z}\gamma^--\beta_{+z}\gamma^+-2\partial_{z}(b_z
c^z ).
\label{ghrepr}
\end{eqnarray}
The ghost contribution to the central charge is $c_{gh}=-6$,
so that $c_{tot}=c_{grav}+c_{gh}=0$, as claimed.

Notice that $\beta$ and $\gamma$ commute among themselves, but
anticommute with $b$ and $c$. This is because they carry an odd ghost number
together with an odd fermion number, while $b$ and $c$ carry
zero fermion number and odd ghost number.

The ghost number charge is
\begin{equation}
{\cal Q}_{gh}=\oint b_{zz}c^z+\beta_{+z}\gamma^++\beta_{-z}\gamma^-+b_zc,
\end{equation}
so that ${\cal Q}_{BRST}=\oint J_z^{BRST}$
has ghost number one:
\begin{equation}
[{\cal Q}_{gh},{\cal Q}_{BRST}]={\cal Q}_{BRST}.
\label{exa}
\end{equation}

We now perform the topological twist on the N=2
gauge-fixed theory.

In order to produce a twisted energy-momentum tensor equal to
$T_{zz}+{1\over 2}\partial_{z} J_z$ we can make
a redefinition of the ghost $c$ of the form
\begin{equation}
c^\prime=c-\partial_{z} c^z.
\label{redef1}
\end{equation}
Such a replacement, which changes the spin of the fields,
is to be viewed as a
redefinition of the $U(1)^\prime$ ghost $C^\prime$ rather
than the $U(1)$ ghost $C$, since the new spin is defined
by adding
the $U(1)^\prime$ charge
(not the $U(1)$ charge) to the old spin, as shown in section
\ref{N2D2}.
$c^\prime$ has a nonvanishing operator product expansion with
$b_{zz}$ so that it is also necessary to redefine $b_{zz}$, namely
\begin{equation}
b_{zz}^\prime=b_{zz}-\partial_{z} b_{z}.
\label{redef2}
\end{equation}
Then, the operator product expansions of the redefined fields are
the same as those for the initial fields.

The spin changes justify
the following change in notation
\begin{equation}
\begin{array}{llll}
\eta_z^+\rightarrow\eta_z,\quad &\quad\lambda_+\rightarrow \lambda,
\quad &\quad \beta_{+z}\rightarrow \beta_z,\quad &\quad
\gamma^+\rightarrow \gamma,\\
\eta^-_{z}\rightarrow\eta,\quad &\quad\lambda_-\rightarrow \lambda_z,
\quad &\quad \beta_{-z}\rightarrow \beta_{zz},\quad &\quad
\gamma^-\rightarrow \gamma^z.
\end{array}
\end{equation}
Similarly, the supercurrents are changed as
$G_{+z}\rightarrow G_z,\quad G_{-z}\rightarrow G_{zz}$.

Redefinitions (\ref{redef1}) and (\ref{redef2}) produce a new
BRST current $J_z^{\prime\, BRST}$ (equal to the old one apart from a
total derivative term) given by
\begin{equation}
J^{\prime\, BRST}_z=-c^z{\cal T}_{zz}^\prime
+{1\over 2}c^\prime {\cal J}_{z}
+{1\over 2}\gamma{\cal G}_{z}-{1\over 2}\gamma^z{\cal G}_{zz},
\end{equation}
where ${\cal T}_{zz}^\prime={\cal T}_{zz}+{1\over 2}\partial_z {\cal J}_z$.
As anticipated,
$J^{\prime\, BRST}_z$ generates
a new energy-momentum tensor (obtained by acting
with the new BRST charge on $b_{zz}^\prime$), which is
\begin{eqnarray}
T_{zz}^\prime&=&T_{zz}+{1\over 2}\partial_{z} J_z
=-\partial_{z} \pi\partial_{z}\varphi+{1\over 2}\partial_{z}^2\pi+
\partial_{z}\chi\partial_{z}\xi+{1\over 2}\partial_{z}^2 \chi-\lambda_z
\partial_{z}\eta
-\partial_{z}\lambda \eta_z
\nonumber\\&&+2b^\prime_{zz}\partial_{z} c^z+\partial_{z} b_{zz}^\prime c^z
+2\beta_{zz}\partial_{z}\gamma^z+\partial_{z}\beta_{zz}\gamma^z
+\beta_{z}\partial_{z}\gamma+b_{z}\partial_{z} c^\prime.
\end{eqnarray}
{}From this expression, it is immediate to check the new spin assignments.
It is interesting to note that the total derivative term in
the $U(1)$ current $J_z^{gh}$ (\ref{ghrepr})
combines with redefinitions (\ref{redef1})
and (\ref{redef2}) to give the correct energy-momentum tensor
for the ghosts $T_{zz}^{\prime\, gh}$.

The other ingredient of the topological twist is the topological shift
\cite{ansfre}
\begin{equation}
\gamma\rightarrow \gamma+\alpha.
\label{redef3}
\end{equation}
Since the spin has been already changed by
(\ref{redef1}), (\ref{redef3}) does not change the spin
a second time. Indeed, the new spin of
$\gamma$ is zero and so that of $\alpha$.
Moreover, after twist
$\gamma$ possesses a zero mode (the constant).
In this case, $\alpha$
represents a shift of the zero mode of $\gamma$.

The topological shift (\ref{redef3})
produces a total BRST current equal to
\begin{equation}
J_z^{BRST\, tot}=J_z^{\prime BRST}+{1\over 2}\alpha {G}_{z},
\end{equation}
(again, a total derivative term has been omitted).
If we denote, as usual,
${\cal Q}_{BRST}=\oint J_z^{BRST\, tot}$, ${\cal Q}_v
=\oint J_z^{\prime\, BRST}dz$ and ${\cal Q}_s=\oint G_{z}dz$,
we see that the BRST charge is precisely shifted by the supersymmetry
charge $Q_s$.

Let us now discuss some properties of the twisted theory.
It is convenient to write down the
${\cal Q}_s$ transformation of the fields, that
we denote it by $\delta_s$:
\begin{equation}
\matrix{
\delta_s(\xi-\varphi)=2\eta,&
\delta_s\eta=0,&
\delta_s\lambda_z=\partial_{z}(\pi+\chi), &
\delta_s(\pi+\chi)=0,\cr
\delta_s(\pi-\chi)=2\lambda,&
\delta_s\lambda=0,&
\delta_s\eta_z=\partial_{z} (\xi+\varphi),&
\delta_s(\xi+\varphi)=0,\cr
\delta_s b_{zz}^\prime=0,&
\delta_s \beta_{zz}=-b_{zz}^\prime,&
\delta_s c^z=\gamma^z,&
\delta_s\gamma^z=0,\cr
\delta_s b_{z}=\beta_{z},&
\delta_s\beta_{z}=0,&
\delta_s c^\prime=0,&
\delta_s \gamma=c^\prime.}
\label{ation}
\end{equation}
These transformations are the analogue, in the gauge-fixed case, of
the $\delta_T$ transformations (\ref{deltaT}) and (\ref{deltaT2}).
Notice that, in the last two lines of (\ref{ation})
there are two different
$b$-$c$-$\beta$-$\gamma$ systems.
In particular, the last line
represents the sector of ${\cal B}_{gauge-fixing}$ that
is reminiscent of the constraint on the moduli space.
The last but one line, on the other hand,
represents the usual $b$-$c$-$\beta$-$\gamma$
ghost for ghost system
of topological gravity \cite{verlindesquare}.
It is evident
that the roles of $b$ and $\beta$ and the roles of $c$
and $\gamma$ are inverted in the two cases.

The theory is topological, since the energy-momentum tensor $T_{zz}^\prime$
is a physically trivial left moving operator.
Indeed, recalling that $G_{zz}=-2\{{\cal Q}_v,
\beta_{zz}\}$, we have
\begin{equation}
\alpha  T_{zz}^\prime=\{{\cal Q}, G_{zz}\}, \quad \quad \{{\cal Q}_v,
G_{zz}\}=0.
\end{equation}

Finally we notice that the ghost number current of the twisted theory
can be written as the sum
of the ghost number charge of the initial N=2 theory plus the $U(1)$
charge. This corresponds to eq.\ (\ref{spina}):
\begin{equation}
{\cal Q}_{gh}^\prime={\cal Q}_{gh}+\oint J_z=
\oint b_{zz}^\prime c^z+2\beta_{zz}\gamma^z+b_zc^\prime-\lambda_z
\eta
-\lambda\eta_z.
\end{equation}

\section{Geometrical Interpretation}
\label{geometry}

We now discuss the moduli space of the twisted theory
and the gauge-fixing sector that implements
the constraint defining the submanifold ${\cal V}_g\subset {\cal M}_g$.

The number of moduli of the twisted theory is $4g-3$,
the same as that of
the N=2 theory, $3(g-1)$ moduli $ m_i $ corresponding to the metric
and $g$ moduli $\nu_j$
corresponding to the $U(1)$ connection $A$.
The number of supermoduli,
on the other hand, changes by one: it was $4(g-1)$ for the N=2
theory, it is $4g-3$ for the topological theory, $3(g-1)$ supermoduli
$\hat  m_i $
corresponding
to the zero modes of the spin 2 antighost $\beta_{zz}$ and $g$
supermoduli $\hat \nu_j$
corresponding to the zero modes of $\beta_{z}$.
The mismatch of one supermodulus is filled by the presence of
one super Killing vector field, corresponding to the (constant)
zero mode of $\gamma$.

In particular, after the twist,
the number of bosonic moduli equals the number of fermionic moduli,
as expected for a topological theory.
However, the two kinds of supermoduli $\hat  m $ and $\hat \nu$
do not carry the same ghost number
after the twist. Indeed, $\hat  m_i $ carry ghost number $1$, while
$\hat \nu_j$ carry ghost number $-1$.
Thus, we can interpret $\hat  m_i $ as the topological variation
of $ m_i $, but we cannot interpret $\hat \nu_j$ as the topological
variation of $\nu_j$, rather $\nu_j$ is the topological
variation of $\hat \nu_j$:
\begin{equation}
\matrix{
s m_i =\hat  m_i ,&\quad s\hat  m_i =0,&
\quad s\hat \nu_j=\nu_j,&\quad s \nu_j=0.}
\label{smu}
\end{equation}
This is in agreement with the interpretation of $A$ as a Lagrange
multiplier, so that it is only introduced {\sl via} the gauge-fixing algebra:
$ m $ and $\hat  m $ belong to ${\cal B}_{gauge-free}$, while
$\nu$ and $\hat \nu$ belong to ${\cal B}_{gauge-fixing}$.

The amplitudes can be written as
\begin{equation}
<\prod_k\sigma_{n_k}>=\int d\Phi
\int_{{\cal M}_g}\prod_{i=1}^{3g-3}d m_i
\int_{{{\bf C}^g / \Lambda}}\prod_{j=1}^gd\nu_j
\int d\hat m  d\hat\nu
\,\prod_i{\rm e}^{q_i\tilde\pi(z_i)}\,
{\rm e}^{-S( m ,\hat m ,\nu,\hat\nu)}\prod_k\sigma_{n_k},
\label{ampltw}
\end{equation}
where $\sigma_{n_k}$ are the observables.
In this expression, the insertions that remove the zero modes of
$b_{zz}$, $\beta_{zz}$, $\beta_{z}$, $b_{z}$, $\eta$, $\lambda$, $\lambda_z$
and $\eta_z$
are understood, but attention has to be paid to the fact that a
super-Killing-vector-field, corresponding to the zero mode of $\gamma$,
forbids one fermionic integration. $e^{q_i \tilde \pi (z_i)}$ are the
$\delta$-type insertions that simulate the curvature R such that
$\sum_i q_i =2(1-g)$, where $\tilde \pi$ is the BRST invariant extension
of $\pi$ \cite{preprint}.
The ghost number of the supermoduli measure
adds up to $-2g+3$. Nevertheless, due to the presence
of one super-vector-field,
the selection rule is that the total ghost number
of $\prod_k\sigma_{n_k}$ must be equal to $2(g-1)$ and not
to $2g-3$. This is the mismatch between true dimension and formal
dimension addressed in the introduction.

To explain why the graviphoton is responsible for the
constraints, let us rewrite the action making the dependence
on the $U(1)$-moduli $\nu_j$ and the corresponding
supermoduli $\hat \nu_j$ explicit.
\begin{eqnarray}
S( m ,\hat m ,\nu,\hat\nu)&=&S( m ,\hat m ,0,0)
+\nu_j\int_{\Sigma_g}\omega^{j}_{\bar z}J_{z}d^2z+
\hat\nu_j\int_{\Sigma_g}\omega^{j}_{\bar z}G_{z}d^2z\nonumber\\&&+
\bar\nu_j\int_{\Sigma_g}\omega^{j}_{z}J_{\bar z}d^2z+
\hat{\bar\nu}_j\int_{\Sigma_g}\omega^{j}_{z}G_{\bar z}d^2z
+\nu\-\hat\nu{\rm -terms}.
\label{actio}
\end{eqnarray}
The terms that are quadratic in $\nu\-\hat\nu$ are due
to the fact that the gravitini are initially $U(1)$-charged. They have
not been reported explicitly, since they can be neglected, as we show
in a moment.
The coefficient of $\bar\nu_j$ is the $U(1)$ current $J_z$
folded with the $j$-th (anti)holomorphic differential $\omega^j_{\bar z}$.
Similarly, the coefficient of ${\hat {\bar\nu}}_j$ is the supercurrent
$G_z$ folded with the same differential.

We want to perform the $\nu$-$\hat\nu$ integrals explicitly.
This is allowed, since the observables should not depend
on  $\nu$ and $\hat\nu$. Indeed,  $\nu$ and $\hat\nu$
belong to ${\cal B}_{gauge-fixing}$ and not to ${\cal B}_{gauge-free}$,
while the observables are constructed entirely
from ${\cal B}_{gauge-free}$. Anyway, since $\nu$ and $\hat\nu$
form a closed BRST subsystem, we can consistently
project down to the subset
$\nu=\hat\nu=0$, while retaining the BRST nilpotence.
The $U(1)$ moduli $\nu$ are not integrated all over ${\bf C}^g$,
which would be nice since the integration would be very easy, rather
on the unit cell $L={\bf C}^g/({\bf Z}^g+\Omega {\bf Z}^g)$
defined by the period matrix $\Omega$. To overcome this problem,
we take the semiclassical limit, which is exact in a topological
field theory. We multiply the action $S$ by a constant $\kappa$
that has to be stretched to infinity. $\kappa$ can
be viewed as a gauge-fixing
parameter, rescaling the gauge-fermion: no physical amplitude depends
on it.
Let us define $\nu^\prime_j=\kappa\nu_j\quad \hat \nu^\prime_j=
\kappa\hat\nu_j$. We have
\begin{equation}
\int_L\prod_{j=1}^gd\nu_jd\hat\nu_j=
\int_{\kappa L}\prod_{j=1}^gd\nu^\prime_jd\hat\nu_j,
\label{p1}
\end{equation}
where  and $\kappa L$ is unit cell rescaled.
We see that the $\nu$-$\hat\nu$-terms of (\ref{actio}) are
suppressed in the $\kappa\rightarrow\infty$ limit, as claimed.
We can replace $\kappa L$ with ${\bf C}^g$ in this limit.
Finally, the integration over the $U(1)$ moduli and supermoduli
produces the insertions
\begin{equation}
\prod_{j=1}^g\int_{\Sigma_g} \omega^j_{\bar z}G_zd^2z\cdot\delta\left(
\int_{\Sigma_g}\omega^j_{\bar z}J_zd^2z\right).
\label{2.47}
\end{equation}

The delta-function is the origin of the desired constraint on moduli space.
Indeed, the current $J_z$ can be thought as a (field dependent)
section of ${\cal E}_{hol}$. The requirement of its vanishing
is equivalent to projecting onto the Poincar\'e dual of the top Chern
class $c_g({\cal E}_{hol})$
of ${\cal E}_{hol}$ \cite{griffithsharris}.
Changing section only changes the representative
in the cohomology class of $c_g({\cal E}_{hol})$.
Indeed, the Poincar\`e dual of the top Chern class of a holomorphic
vector bundle $E\rightarrow M$ is shown to be the submanifold of the base
manifold $M$ where one holomorphic section $a\in \Gamma(E,M)$
vanishes identically. In other words, the dual of $c_{g}({\cal
E}_{hol})$ is the divisor of some section. For a line bundle
$L\rightarrow M$, this is easily seen. Let $h$ be a fiber metric so
that
$||a||^2=a(z)\bar a(\bar z)h(z,\bar z)$ is the norm of the
section $a$. The top Chern class $c_1(L)$ can be written as the
cohomology class of the curvature $R=\bar \partial\Gamma$
of the canonical holomorphic connection $\Gamma=h^{-1}\partial h$,
so that $c_1(L)=\bar \partial\partial\ln ||a(z)||^2$.
Patchwise, the metric $h$ can be reduced to the identity, but then
$c_1(L)$ becomes a de Rham current, namely a singular $(1,1)$ form
with delta-function support on the divisor Div[$a$], i.e.\ the
locus of zeroes and poles of $a(z)$. The divisor Div[$a$]
is the Poincar\`e dual of $c_1(L)$.
For a holomorphic vector bundle $E\rightarrow M$ of rank $n$, the
same theorem can be understood using the so-called splitting principle
, regarding $E$ as the Whitney sum of n line-bundles $L_i$
corresponding naively, to the eigendirections of the curvature matrix two form
${\cal R}^{jk}$.
{}From the above argument, we can say that $c_n(E)$ has delta-function support
on the divisor of $a$.
That is why in our derivation of the topological correlators from the
functional integral, we do not pay particular
attention to the explicit form of $J_z$ and to its dependence on the
other fields. What matters is that it is a conserved holomorphic one
form, namely a section of ${\cal E}_{hol}$. The functional integral
imposes its vanishing, so that the Riemann surfaces
that effectively contribute lie in the homology class of the
Poincar\`e dual of $c_g({\cal E}_{hol})$.

Summarizing, we argue that the topological observables $\sigma_{n_k}$
correspond to the Mum\-ford-Morita classes, as in the
case of topological gravity \cite{verlindesquare},
but that in constrained topological gravity the
correlation functions are intersection forms
on the Poincar\'e dual ${\cal V}_g$ of $c_g({\cal E}_{hol})$ and not on the
whole moduli space ${\cal M}_g$.

It can be convenient to
represent
$c_g({\cal E}_{hol})$ by introducing the natural
fiber metric $h_{jk}={\rm Im}\, \Omega^{jk}=\int_{\Sigma_g}
\omega^j_z\omega^k_{\bar z}d^2z$ on ${\cal E}_{hol}$.
The canonical connection associated
with this metric is then $\Gamma=
h^{-1}\partial h={1\over \Omega-\bar \Omega}\partial \Omega$. Then
${\cal R}=\bar\partial \Gamma$ and
\begin{equation}
c_g({\cal E}_{hol})=\det {\cal R}=\det\left({1\over \Omega-\bar \Omega}
\bar \partial\bar \Omega{1\over \Omega-\bar \Omega}
\partial\Omega\right).
\end{equation}

Let $\{\omega^1,\ldots \omega^g\}$
denote a basis of holomorphic differentials.
Locally, we can expand $J_z$ in this basis $J_z=a_j\omega^j_z$.
The field dependent
coefficients $a_j$ are the components of the section $J_z\in\Gamma(
{\cal E}_{hol},{\cal M}_g)$.
The constraint then reads ${\rm Im}\,\Omega^{jk}a_k=0,\ \forall
j$, which, due to the positive definiteness of ${\rm Im}\,\Omega$, is
equivalent
to
\begin{equation}
a_j=0,\quad\quad\forall j.
\end{equation}
These are the equations that (locally) identify the
submanifold ${\cal V}_g\subset {\cal M}_g$.
It is also useful to introduce
the vectors $v_j={\partial\over \partial a_j}$ that
provide a local basis for the normal bundle ${\cal N}({\cal V}_g)$ to
${\cal V}_g$. Of course, the vectors $v_j$ commute among themselves:
$[v_j,v_k]=0$.
In these explicit local coordinates, the top Chern class
$c_g({\cal E}_{hol})$ admits the following representation as a
de Rham current:
\begin{equation}
c_g({\cal E}_{hol})=\delta({\cal V}_g)\tilde\Omega_g,
\label{9.15}
\end{equation}
where
\begin{equation}
\tilde\Omega_g=\prod_{j=1}^g da_j,\quad\quad
\delta({\cal V}_g)=\prod_{j=1}^g\delta (a_j).
\end{equation}
This explicit notation is useful to trace back the correspondence
between the geometrical and field theoretical definition of the
correlators.

To begin with,
a convenient representation of the BRST operator (\ref{smu})
on the space
$\{ m ,\hat  m ,\nu,\hat\nu\}$ is given by
\begin{equation}
{\cal Q}_{global}=\hat  m_i {\partial\over \partial  m_i }+
\nu_j{\partial\over \partial \hat \nu_j}.
\label{operator}
\end{equation}
${\cal Q}_{global}$
is not the total BRST charge, rather it only represents the
BRST charge on the sector of the
global degrees of freedom. The total BRST
charge is the sum of the above operator plus the usual BRST charge
${\cal Q}={\cal Q}_s+{\cal Q}_v$,
that acts only on the local degrees of freedom.
Since the total BRST charge acts trivially inside the physical
correlation functions, we see that the action of ${\cal Q}$
inside correlation
functions is the opposite of the action of ${\cal Q}_{global}$.
This means that
${\cal Q}$ can be identified, apart from an overall immaterial sign,
with the operator (\ref{operator}).
We know that the geometrical meaning of the supermoduli $\hat m_i$
are the differentials $d m_i $ on the moduli space ${\cal M}_g$.
The ghost number corresponds to the form degree. In view of this, we
argue that the geometrical meaning of the $U(1)$ supermoduli
$\hat \nu_j$ are contraction operators
$i_{v_j}$ with respect to the associated vectors $v_j$.
Since the $U(1)$ moduli $\nu_j$ are the BRST variations of $\hat\nu_j$
and the BRST operation should be
identified with the exterior derivative,
it is natural to conjecture that $\nu_j$ correspond to
the Lie derivatives along the vectors $v_j$.

The correspondence between field theory and geometry is summarized
in table \ref{table1}.
We now give arguments in support of this interpretation.

For instance, since ${\cal Q}\sim d$,
${\rm Im}\,\Omega^{jk}\, a_k\sim \int\omega_{\bar z}^kJ_zd^2z$ and
$[{\cal Q},J_z]=-G_z$, then
the insertions $\int \omega_{\bar z}^jG_zd^2z$ correspond to
$d({\rm Im}\,\Omega^{jk}a_k)$, so that
\begin{equation}
\prod_{j=1}^g\int_{\Sigma_g} \omega^j_{\bar z}G_zd^2z\cdot\delta\left(
\int_{\Sigma_g}\omega^j_{\bar z}J_zd^2z\right)\sim
\tilde\Omega_g\delta({\cal V}_g)=c_g({\cal E}_{hol}).
\end{equation}

If $\alpha_k$ denote the Mumford-Morita classes corresponding
to the observables ${\cal O}_k$, the amplitudes are
\begin{equation}
<{\cal O}_1\cdots {\cal O}_n>=
\int_{{\cal M}_g}\delta({\cal V}_g)\tilde\Omega_g\wedge \alpha_1
\wedge\cdots\wedge \alpha_n=\int_{{\cal V}_g}\alpha_1
\wedge\cdots\wedge \alpha_n.
\end{equation}

{}From the geometrical point of view, it is immediate to
show that the action of (\ref{operator}) on a correlation
function is precisely the exterior derivative, as already advocated.
Indeed, we can write the $d$-form $\omega_d$ corresponding to a
physical amplitude (not necessarily a top form, if we freeze, for the moment,
the integration over the global degrees of freedom) as
$\omega_d=i_{v_1}\cdots i_{v_g}\Omega_{d+g}=
\left(\prod_{j=1}^g\hat\nu_j\right)\hat m_{i_1}\cdots\hat m_{i_d}
\Omega_{d+g}^{i_1\cdots i_d}$,
where $\Omega_{d+g}$ is a suitable $d+g$-form on ${\cal M}_g$
(equal to $\tilde\Omega_g\wedge\omega_d$).
Now, using the representation (\ref{operator}) of the operator
${\cal Q}$ and the correspondence given in table II we find precisely
$\{{\cal Q},\omega_d\}=d \omega_d$.
The second piece of (\ref{operator}) replaces a
contraction with the vector
$v_j$ with the Lie derivative with respect to the same vector.

\vfill\eject

\vfill\eject

\begin{table}
\begin{center}
\caption{Field Theory versus Geometry}
\begin{tabular}{|c|c|}
\hline
Field Theory & Geometry\quad \\
\hline
\quad $\hat  m_i $     & $d m_i $\quad \\
\quad $\hat \nu_j$     & $i_{v_j}$\quad \\
\quad $\nu_j$          & ${\cal L}_{v_j}$\quad \\
\quad ${\cal Q}$       & $d$\quad \\
\quad $[{\cal Q}, m_i ]=\hat m_i$    & $[d, m_i ]=d m_i $\quad \\
\quad $\{{\cal Q},\hat m_i \}=0$      & $\{d,d m_i \}=0$\quad \\
\quad $\{{\cal Q},\hat \nu_j\}=\nu_j$ & $\{d,i_{v_j}\}={\cal L}_{v_j}$\quad \\
\quad $[{\cal Q},\nu_j]=0$            & $[d,{\cal L}_{v_j}]=0$\quad \\
\quad $\{\hat\nu_j,\hat\nu_k\}=0$     & $\{i_{v_j},i_{v_k}\}=0$\quad \\
\quad $[\hat\nu_j,\nu_k]=0$           & $[i_{v_j},{\cal L}_{v_k}]=0$\quad \\
\quad $[\nu_j,\nu_k]=0$               & $[{\cal L}_{v_j},{\cal L}_{v_k}]=0$
\quad \\
\quad $\prod_{j=1}^g \delta\left(\int \omega^j_{\bar z}
J_zd^2z\right)$&
$\delta({{\cal V}_g})$\quad\\
\quad $\prod_{j=1}^g \int\omega^j_{\bar z}
G_zd^2z$ &$\tilde\Omega_g$\quad\\
\quad  $\prod_{j=1}^g \int\omega^j_{\bar z}
G_zd^2z\cdot\delta\left(\int \omega^j_{\bar z}J_zd^2z\right)$
& $c_g({\cal E}_{hol})=\delta({{\cal V}_g})\tilde\Omega_g$\quad\\
\quad $\sigma_{n_j}$   & $[c_1({\cal L}_j)]^{n_j}$\quad\\
\quad $<\sigma_{n_1}\cdots \sigma_{n_k}>$ & $\int_{{\cal V}_g}
[c_1({\cal L}_1)]^{n_1}\wedge\cdots \wedge
[c_1({\cal L}_k)]^{n_k}$\quad \\
\hline
\end{tabular}
\end{center}
\label{table1}
\end{table}

\end{document}